\documentclass[preprint,3p]{elsarticle}

 \usepackage{graphicx}

\usepackage{amssymb}
\usepackage{bm}

\usepackage{amsmath}
\journal{Physics Letters A}
\usepackage{color}

\begin{document}

\begin{frontmatter}



\title{Magnetization and Lyapunov exponents on a kagome chain with multi-site
 exchange interaction}


\author[erphi]{N. Ananikian}
\author[erphi]{L. Ananikyan}
\author[udsd]{R. Artuso}
\author[ysu]{H. Lazaryan}

\address[erphi]{Yerevan Physics Institute, Alikhanian Br. 2, 0036
Yerevan, Armenia,}
\address[udsd]{Dipartimento di  Fisica e Matematica, Universita'
degli Studi dell'Insubria,\\ Via Valleggio 11, 22100
Como, Italy,\\I.N.F.N. Sezione di Milano, Via Celoria 16, 20132 Milano, Italy}
\address[ysu]{Department of Theoretical Physics, Yerevan State University,\\ A. Manoogian 1, 0025 Yerevan, Armenia,}

\begin{abstract}
The Ising approximation of the Heisenberg model in a strong magnetic field, with two, three and six spin exchange interactions is studied on a kagome chain. The kagome chain can be considered as an approximation of the third layer of $^3He$  absorbed on the surface of graphite (kagome lattice). By using dynamical approach we have found one and multi-dimensional mappings (recursion relations) for the partition function. The magnetization diagrams are plotted and they show that the kagome chain is separating into four sublattices with different magnetizations. Magnetization curves of two sublattices exhibit plateaus at zero and 2/3 of the saturation field. The maximal Lyapunov exponent for multi-dimensional mapping is considered and it is shown that near the magnetization plateaus the maximal Lyapunov exponent also exhibits plateaus.  \end{abstract}


\end{frontmatter}


\section{Introduction}
 The partition function of a statistical system can be calculated by means of several methods: transfer matrix approach \cite{books}, dynamic technique \cite{obzordynamic1}-\cite{dynamicheski2} etc.  The most useful for recursive lattices is the dynamical approach, which is a very powerful tool for research of a number of theoretical problems in statistical mechanics. By means of this approach one can obtain exact recursive relations for the partition function on recurrent lattices.

As the systems of almost localized identical fermions one can model solid  and fluid $^3He$ films. Since the light mass spin-1/2  $^3He$
 atoms are subject to a week attractive potential, the theoretical explanation of magnetism is based on the multiple-spin exchange mechanism \cite{Roger}.
An important case is represented by
solid and fluid $^3He$ films absorbed on
the surface of
graphite \cite{graffite,graffite2,graffite3} since it is a
typical example of a two-dimensional frustrated quantum-spin
system \cite{frustrated}. The first and second layers of
the system form a triangular lattice, while the third one forms a
system of quantum 1/2 spins on a kagome
lattice \cite{kagome2}.

In a strong external magnetic field   Heisenberg model can be
approximated by  the Ising  one. If the strong magnetic field is directed along the z-axis, a reduction of transverse fluctuations is expected to occur; it is supposed that in this case  $\sigma^x$ and
$\sigma^y$-spin components are infinitely small infinitely small and may be neglected  \cite{Roger,platoes6}.

The phenomenon of magnetization plateau has been studied during the past decade both experimentally
and theoretically. The plateaus may be exhibited in the magnetization curves of quantum spin systems at very low temperatures in case of strong external fields.
 The plateaus of this kind were first observed by   Hida \cite{hida} in trimerized S=1/2 spin
chains and they were analytically explained by Okamoto \cite{okamato}.
Magnetization
plateaus appear in a wide range of  models on chains, ladders,
hierarchical lattices, theoretically analysed by
dynamical, transfer matrix
approaches and  exact diagonalization in clusters (see Ref.~\cite{platoes1}-\cite{platoes13}). The experimental evidence for the appearance of magnetization plateaus has also been found for two-dimensional systems \cite{experemental}-\cite{experementa5}. On triangular lattices a magnetization
plateau was observed at $m/m_s=1/3$  for compounds like
C$_6$Eu \cite{CEu1},  CsCuCl$_3$ \cite{CsCuCl} (see also Ref.
\cite{NH4CuCl32}-\cite{NH4CuCl36}); plateaus have also been observed on a kagome lattice  (see Ref. \cite{kagomeplato}-\cite{kagomeplato12}).

In the present paper the dynamical (recursive) approach was used to study magnetic properties of a kagome chain. The magnetization plateaus were found at low temperatures and moreover, the kagome chain was separated into four sublattices with different magnetizations. Two of these exhibited plateaus, whereas the others did not.  The stability
of the system is described by Lyapunov exponents \cite{lyapunov3} by means of which the exponential rate is measured, at which the adjacent  orbits converge or diverge \cite{lyapunov1}-\cite{lyapunov6}.
It is interesting to check whether the maximum Lyapunov exponent has plateaus that coincide with the magnetization one.

This paper is organized as follows: in Section 2 the two, three and six-site exchange interactions in a strong magnetic field are described, in Section 3 the exact recursion relation for the partition function is derived, the magnetization curves for different values of temperature and exchange parameters are displayed in Section 4. In Section 5 a multidimensional mapping for the partition function is obtained and the corresponding maximum Lyapunov exponent is investigated. Finally, Section 6 contains the concluding remarks.

\section{Exchange Hamiltonian with two, three and six- site exchange }
The Heisenberg Hamiltonian for kagome chain consists of two parts \begin{equation}
H = H_{ex}  + H_Z,
\end{equation}
where $H_{ex}$ is the spin exchange Hamiltonian and $H_Z$ is the
Zeeman Hamiltonian which is responsible for magnetism, which has the following
form
\begin{equation}\label{Hz}
H_Z  =  - \sum\limits_i {\frac{\gamma }{2}\hbar \bm{B}\bm{\sigma} _i}
=-\mathrm{h}\ \sum\limits_i {\bm{\sigma} _i}
,\end{equation}
where $\gamma$  is the gyromagnetic ratio for a nucleus, and $\bf B$  is the magnetic field \cite{Roger}.

 According to \cite{Roger} the multiple spin exchanges Hamiltonian  can be
written as
\begin{equation}
\label{Hamiltonian23}
H_{ex}  = \mathrm{J}_2 \sum\limits_{Pairs} {P_2 }  - \mathrm{J}_3 \sum\limits_{Triangles}
{\left( {P_{3}  + P_{3}^{ - 1} } \right)+}
 \mathrm{J}_6 \sum\limits_{Hexagons}
{\left( {P_{6}  + P_{6}^{ - 1} } \right)},
\end{equation}
where $P_{2}$ is the pair transposition operator, $P_{3}$ ($P_{6}$) is the operator making a cyclic rearrangement in the triangle (hexagon). The explicit expression
of pair transposition operator $P_{ij}$ was been given by Dirac\begin{equation}\label{p2}
 P_{2}\equiv P_{ij}  = \frac{1}{2}\left( {1 +  \bm{\sigma} _i
\bm{\sigma} _j } \right) ,\end{equation} where $\bm{\sigma} _i$ is
the Pauli matrix, acting on the spin at the $i$-th site. The expressions for $P_3$, and $P_6$ in terms of Pauli matrixes were derived in \cite{platoes11}. By using these expressions one can rewrite
the Heisenberg Hamiltonian in terms of Pauli matrices. By taking into account that in a strong external magnetic field directed along the z-axis the contribution from x and y spin component will be negligible and the only relevant contribution will come from the z component (which can effectively take values $s_z = \pm1$),
 we may consider an Ising model instead of  Heisenberg one. By using expressions for $P_3$, and $P_6$ and by substituting the Pauli matrix with $s_z$
we get  the Ising approximation of the Heisenberg Hamiltonian:
\begin{equation}\label{HamiltonianIsing}
H = \frac{{\mathrm{J}_2 }}{2}\sum\limits_{Pairs} {\left( {1 + s
_i s _j } \right)-}  \frac{{\mathrm{J}_3
}}{2}\sum\limits_{Triangles} {\left( {1 + s _i s _j  + s _j s _k +
s _k s _i } \right)}+\frac{{\mathrm{J}_6 }}{2}\sum\limits_{Hexagons}{H^{(6)}}
-\mathrm{h}\sum\limits_i {s _i },
\end{equation}
where $H^{(6)}$ represents the six particle exchange term in each hexagon:
\begin{equation}
H^{(6)} = \frac{1}{8}\left( 1 + \sum\limits_{\mu  < \nu } {s _\mu s
_\nu  } + \sum\limits_{\mu  < \nu  < \lambda < \rho } {s _\mu s
_\nu  s _\lambda  s _\rho }+  s _1 s _2 s _5 s _6 s _9 s
_{10} \right)
\end{equation}
where the first sum goes over all pairs
in the hexagon, and the second one goes over all quartets in the hexagon.
\begin{figure}
\center
\includegraphics[width=350pt]{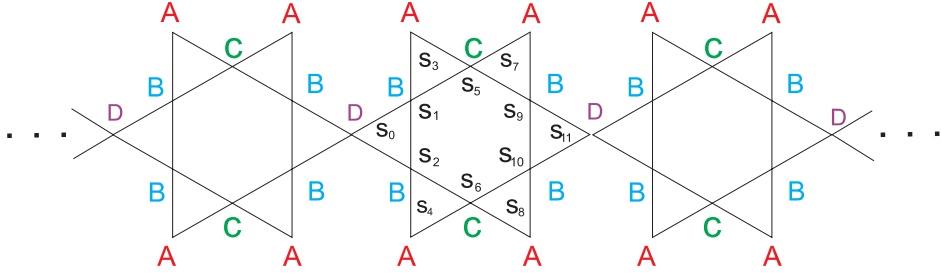}
\caption{Sublattices on the kagome chain.}
\end{figure}
\section{The recursion relation for the partition function}
The partition function of the system
is
defined in the following way\begin{equation}\label{ZZZ}
Z = \sum\limits_{\{ s _i \} } {e^{ -   H\left(s_1,s_2,...s_n
\right)/k_BT}} ,\end{equation} where $   H\left(s_1,s_2,...s_n
\right)$ is the Hamiltonian of the system, $k_B$ is the Boltzmann constant,
$T$ is the temperature  and the sum goes over all $\{s_1,s_2,...s_n\}$
configurations of the system. Below all the parameters will be rescaled using the Boltzmann constant
($\mathrm{J}_2/k_B \equiv J_2,\ \mathrm{J}_3/k_B \equiv J_3,\
\mathrm{h}/k_B \equiv h$).
To obtain recursion relations for the partition function one can separate the kagome ladder into two identical parts (branches) and firstly perform a summation over all spin configurations on each branch, and secondly sum over the central spin variable (see Fig. 1). The summation on each branch yields the same result and such a term only depends on the value of central spin:\begin{equation}\label{partitionfuncton}
Z = \sum\limits_{s _0 } {e^{\frac{h}{T}s _0 }g_n^2 (s _0 )} =
e^\frac{h}{T}g_n^2 ( + ) + e^{ -\frac{ h}{T}}g_n^2 ( - ),
\end{equation}
where $g_n(s _0)$ denotes the contribution of each branch.
 The expression for    $g_n(s_{0})$   can be written ($s_1\ldots s_{11}$, see Fig. 1) in the form:  \begin{equation}\label{grecurent}
g_{n}(s _0 ) = \sum\limits_{s _1 ,s _3 ...s _{11} }
{e^{k(s_{0},...s_{11})}g_{n-1}(s_{11})},
\end{equation}
where and
$n$ is the  number of generations and the expression for $k(s_{0},...s_{11})$ has the following form: \begin{equation}
k(s_{0},...s_{11})=-\frac{{J_2
}}{2T}\sum\limits_{18\ pairs}
 \left( {1 + s _i s _j   } \right)+
\frac{{J_3 }}{2T}\sum\limits_{6\ triangles}
 \left( {1 +
s _i s _j  + s _j s _k  + s _k s _i
} \right) - \frac{{J_6}}{2T}\cdot H^{(6)}  + \frac{h}{T}\sum\limits_{i =1}^{11} {s _i. }
\end{equation}
Since $g_{n-1}$ does not depend on $s_0,...s_{10}$, equation (\ref{grecurent})
may be obtained as \begin{eqnarray}
g_n ( + ) &=& K( + , + )g_{n - 1} ( + ) + K( + , - )g_{n - 1} ( -
),\nonumber\\
 g_n ( - ) &=& K( - , + )g_{n - 1} ( + ) + K( - , - )g_{n - 1} ( -
 ),\nonumber\\
 \end{eqnarray}
 where  we denote $g_n(s_i)$ by $g_n(\pm)$ according to the sign of $s_i$  and
$K(s_0,s_{11})$ is the sum of $k(s_{0},...s_{11})$ over the ten intermediate spins
between $s_1$ and $s_{12}$
\begin{equation}
K(s_1,s_{12})= \sum\limits_{s _2 ,s _3 ...s _{11} }
{e^{k(s_{0},...s_{11})}.}
\end{equation}
The recursion relation is obtained by introducing a new variable
$x_n = g_n ( + )/g_n ( - )$:
\begin{equation}\label{recurentfunction}
x_n = \frac{K( + , + )x_{n - 1} + K( + , - )}{K( - , + )x_{n - 1} + K( - , -
)}.
\end{equation}

 \section{Magnetization plateaus}
 \begin{figure}
\center
\includegraphics[width=\textwidth]{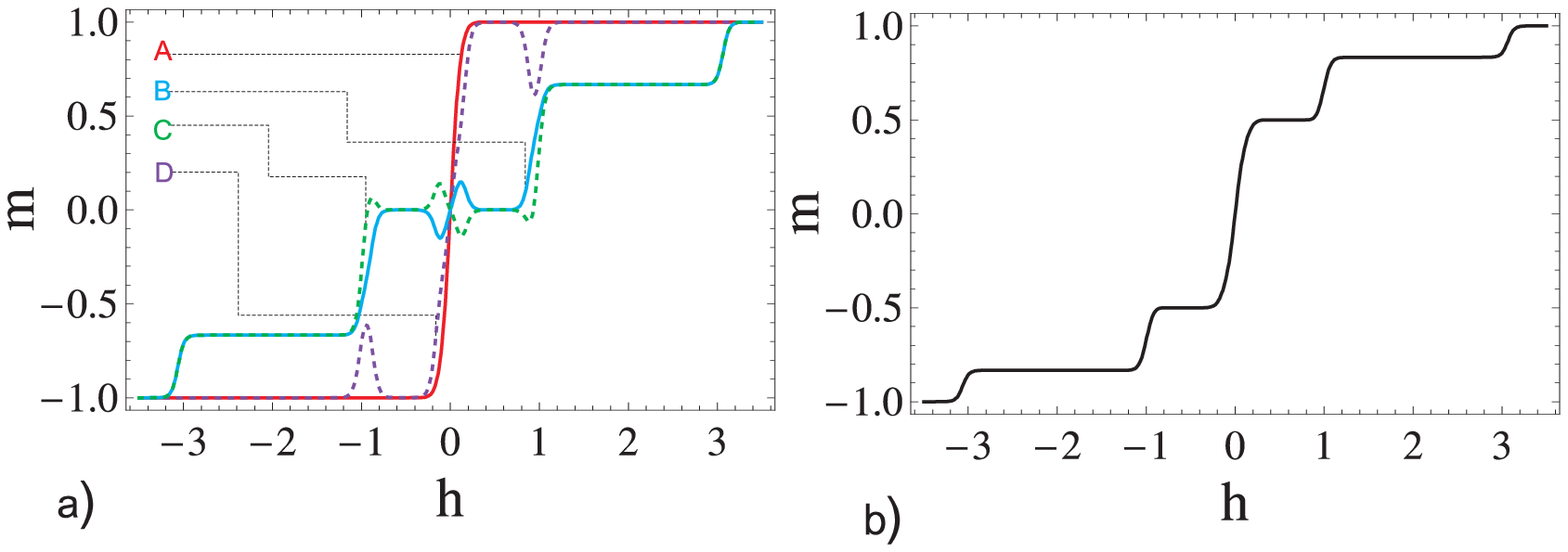}
\caption\ { Magnetization curves for $J_2=2mK,J_3=1.5mK,J_6=2mK,
T=0.04mK $  for a) different sublattices  \\ b) the average magnetization of
 all sublattices }
\end{figure}
 The Thermodynamic functions of the system,
such as magnetization, can be expressed in terms of $x_n$. The local
magnetization at $s^{}_r$ site is expressed
as\begin{equation}\label{magnetization} m=\left\langle
s^{}_r\right\rangle=\frac{P(s_{r})}{Z},
\end{equation}
where
\begin{equation}
P(s_{r})\equiv\sum\limits_{\{ s _i \} } {
s^{}_re^{ - H\left(s_1,s_2,...s_n \right)/k_BT}}.
\end{equation}
By using equation (\ref{partitionfuncton})
one can write for the central vertex\begin{equation}\label{magnit}
m=\frac{{e^{ \frac{h}{T}} g_n^2 ( + ) - e^{ -\frac{  h}{T}} g_n^2 ( - )}}{{e^{ \frac{h}{T}} g_n^2 ( + ) + e^{ -\frac{  h}{T}} g_n^2 ( - )}} = \frac{{e^{ \frac{2h}{T}}
x_n^2  - 1}}{{e^{ \frac{2h}{T}}x_n^2  + 1}}
.\end{equation}
To obtain local magnetization for non central vertices one should replace one of two multipliers    $g_n(s_{1})$  in (\ref{partitionfuncton})
by formula (\ref{grecurent}) (see
Fig. 1) \begin{equation}
Z= \sum\limits_{s _0 ,s _2
...s _{11} } {  e^{ \frac{h}{T}s_0}\cdot e^{ k(s_{0},...s_{11})}\cdot g_{n}(s_{0})\cdot g_{n-1}(s_{11})}.
\end{equation}
Suppose it is required to find the local magnetization for the  $r$-th site ($r\neq0,11$):
in that case we must expand sums over $s_0$, $s_{11}$ and $s_r$
\begin{eqnarray}\label{zrep}
Z&=& \left( K_{ + , + }^+ + K_{ + , + }^-  \right)\cdot g_n ( +
)g_{n-1}(+)+ \left( K_{ - , + }^ +   + K_{ - , + }^ - + K_{ + , - }^ + + K_{ +
, - }^ -   \right)\cdot g_n ( + )g_{n-1} ( - ) +\nonumber\\
&+& \left( K_{ - , - }^ + +
K_{ - , - }^ -   \right)\cdot g_n ( - )g_{n-1} ( - ),
\end{eqnarray}
the sum\begin{equation}
K_{ s_0 , s_{11} }^{s_r}  = \sum\limits_{s _0..s _{r-1},s_{r+1}
...s _{11} } {  e^{ \frac{h}{T}s_0}\cdot e^{ k(s_{0},...s_{11})}}
\end{equation}
is denoted as  $K_{ \pm , \pm }^\pm $ depending on the sign of $s_0,s_{11}$ and $s_r$ sign.
In the same way the  expression for $P(s_r)$ may be obtained.  The difference is that each $K_{ \pm , \pm }^-$ is replaced by $(-K_{ \pm , \pm }^-)$:
\begin{eqnarray}\label{prep}
P(s_r)&=& \left( K_{ + , + }^+ - K_{ + , + }^-  \right)\cdot g_n ( +
)g_{n-1}(+)+ \left( K_{ - , + }^ +   - K_{ - , + }^ - + K_{ + , - }^ + - K_{ +
, - }^ -   \right)\cdot g_n ( +
)g_{n-1}(-) +\nonumber\\
&+& \left( K_{ - , - }^ + -
K_{ - , - }^ -   \right)\cdot g_n (-
)g_{n-1}(-).
\end{eqnarray}
By using equations (\ref{zrep}) and (\ref{prep})
we find the following expression for the magnetization at $r$-th site
in terms of $x_n$

\begin{equation}\label{prep1}
m_{r}=\frac{ \left( K_{ + , + }^+ - K_{ + , + }^-
\right)\cdot x_n x_{n-1} + \left( K_{ - , + }^ +   - K_{ - , + }^ - + K_{ + , - }^ + - K_{ +
, - }^ -   \right)\cdot x_n + \left( K_{ - , - }^ + -
K_{ - , - }^ -   \right)}{\left( K_{ + , + }^+ + K_{ + , + }^-  \right)\cdot
x_n x_{n-1} + \left( K_{ - , + }^ +   + K_{ - , + }^ - + K_{ + , - }^ + + K_{ +
, - }^ -   \right)\cdot x_n  + \left( K_{ - , - }^ + +
K_{ - , - }^ -   \right)}.
\end{equation}

For fixed values of temperature and exchange parameters the dependence of magnetization on the external magnetic field may be obtained by means of a simple iteration scheme induced by the recursion relations, starting from some initial value $x_0$,  The thermodynamic limit then corresponds to the
asymptotic limit $(n\rightarrow\infty)$.

Because of the symmetry of the Hamiltonian, the values of magnetization for some
spins are equal: we have in general (See Fig. 1),\begin{eqnarray}
m(s_3)=m(s_4)=m(s_7)=m(s_{8})&-&A\nonumber\\
m(s_1)=m(s_2)=m(s_9)=m(s_{10})&-&B\nonumber \\
m(s_5)=m(s_6)&-&C\nonumber \\
m(s_0)=m(s_{11})&-&D
\end{eqnarray}

In Figure 2 magnetization curves are plotted for different sites. As it can be
seen from the figure, the magnetization functions have plateaus only for the sublattices B and
C.
\section{Two dimensional mapping and Lyapunov exponent}
\begin{figure}
\center
\includegraphics[width=350pt]{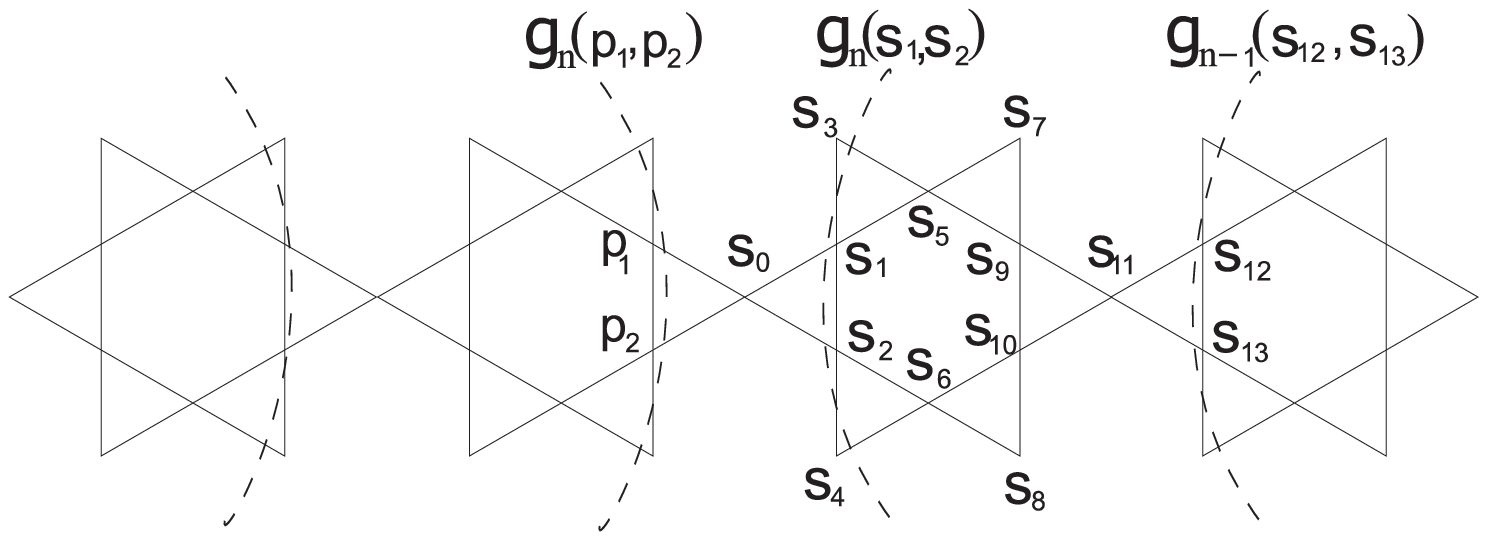}
\caption{The kagome chain.}
\end{figure}
\textcolor{red}{}It is interesting to calculate the Lyapunov exponent near plateaus
but one cannot use recurrent function (\ref{recurentfunction}) because it is
related to $s_1$ site.  To obtain the recursion relation for the partition
function that is related to $s_2$ site one can separate the kagome chain  into two identical
parts as shown in Figure 3:
\begin{equation}\label{Partition2}
Z = \sum\limits_{s_1,s_2,~ p_1,p_2 } {N^{s_1,s_2}_{p_1,p_2}\cdot g_n(s_1,s_2)\cdot g_n(p_1,p_2)},
\end{equation}
where
\begin{equation}\label{Ns1s2p1p2}
 N^{s_1,s_2}_{p_1,p_2}=\sum\limits_{s_0 }e^{\left\{\frac{h}{T}(s_1+s_2+s_0+p_1+p_2)+\frac{J_3 - J_2 }{2T}\left(H^{(2,3)}(s_0,s_1,s_2)+H^{(2,3)}(s_0,p_1,p_2)\right)-\frac{{J_6}}{2T} H^{(6)}\right\}}
\end{equation}and the contribution of each branch is denoted by  $g_n(s _1 ,s_{2})$. The recursion relations for $g_n$  can be obtained by performing
 eleven ($s_3\ldots s_{13}$, see Fig. 3) summations in the kagome chain: \begin{equation}\label{grecurent2}
g_{n}(s _1 ,s_{2}) = \sum\limits_{s_{12},s _{13} }
{K^{s_1,\ s_2}_{s_{12},s_{13}}\cdot g_{n-1}(s_{12},s_{13})},
\end{equation}
where  $ K^{s_1,\ s_2}_{s_{12},s_{13}}$ is as follows
\begin{equation}
K^{s_1,\ s_2}_{s_{12},s_{13}} = \sum\limits_{s_3\cdots s _{11} }e^{\frac{J_3 - J_2 }{2T}\sum\limits_{6\ Triangle}
{ H^{(2,3)}}   -\frac{{J_6}}{2T} H^{(6)}  + \frac{h}{T}\sum\limits_{i =
3}^{13} {s _i }}.
\end{equation}
Since $K^{s_1,\ s_2}_{s_{12},s_{13}}=K^{s_2,\ s_1}_{s_{12},s_{13}} ,K^{s_1,\ s_2}_{s_{12},s_{13}}=K^{s_1,\ s_2}_{s_{13},s_{12}}  $ one can show that  $g_n(+,-)=g_n(-,+)$ and, hance, the recurrence
relations are two-dimensional.
  By
introducing  $ x_n = g_n ( +,+ )/g_n ( -,- ) $ and  $ y_n = g_n ( +,- )/g_n ( -,- )    $  the recursion
relations may be written in the following form:
\begin{eqnarray}\label{rec2}
x_{n}=\Phi(x_{n-1},y_{n-1}),\  \Phi(x,y_{})&=&
\frac{K_{+,+}^{+,+}x+2K_{+,-}^{+,+}y+K_{-,-}^{+,+}}{K_{+,+}^{-,-}x+2K_{+,-}^{-,-}y+K_{-,-}^{-,-}},\nonumber\\
y_{n}=\Psi(x_{n-1},y_{n-1}),\  \Psi(x,y_{})&=&
\frac{K_{+,+}^{+,-}x+2K_{+,-}^{+,-}y+K_{-,-}^{+,-}}{K_{+,+}^{-,-}x+2K_{+,-}^{-,-}y+K_{-,-}^{-,-}}.
\end{eqnarray}

The magnetization can be expressed  in terms of $x_n,y_n$ with due regard for
(\ref{Partition2}) and (\ref{Ns1s2p1p2})
\begin{equation}\label{ms2}
m(s_1) = \frac{\sum\limits_{s_1,s_2,p_1,p_2 } {s_1\cdot N^{s_1,s_2}_{p_1,p_2}\cdot g_n(s_1,s_2)\cdot g_n(p_1,p_2)}}{\sum\limits_{s_1,s_2,p_1,p_2 } { N^{s_1, s_2}_{p_1,p_2}\cdot g_n(s_1,s_2)\cdot g_n(p_1,p_2)}}.
\end{equation}
Taking into account  that $N^{s_1,s_2}_{p_1,p_2}=N^{s_2,s_1}_{p_1,p_2}=N^{s_1,s_2}_{p_2,p_1}=N^{p_1,p_2}_{s_1,s_2}$, we get after the expansion of sums:
\begin{equation}\label{ms2final}
m(s_1) = \frac{N^{+,+}_{+,+}x_n^2+2N^{+,+}_{+,-}x_ny_n-2N^{+,-}_{-,-}y_n-N^{-,-}_{-,-}}{N^{+,+}_{+,+}x_n^2+4N^{+,+}_{+,-}x_ny_n+4N^{+,-}_{+,-}y_n^{2}+2N^{+,+}_{-,-}x_n+4N^{+,-}_{-,-}y_n+N^{-,-}_{-,-}}.
\end{equation}
Magnetization curves calculated by formulae  (\ref{ms2final}) and (\ref{prep1}) with $r=1$
coincide.
\begin{figure}
\center
\includegraphics[width=250pt]{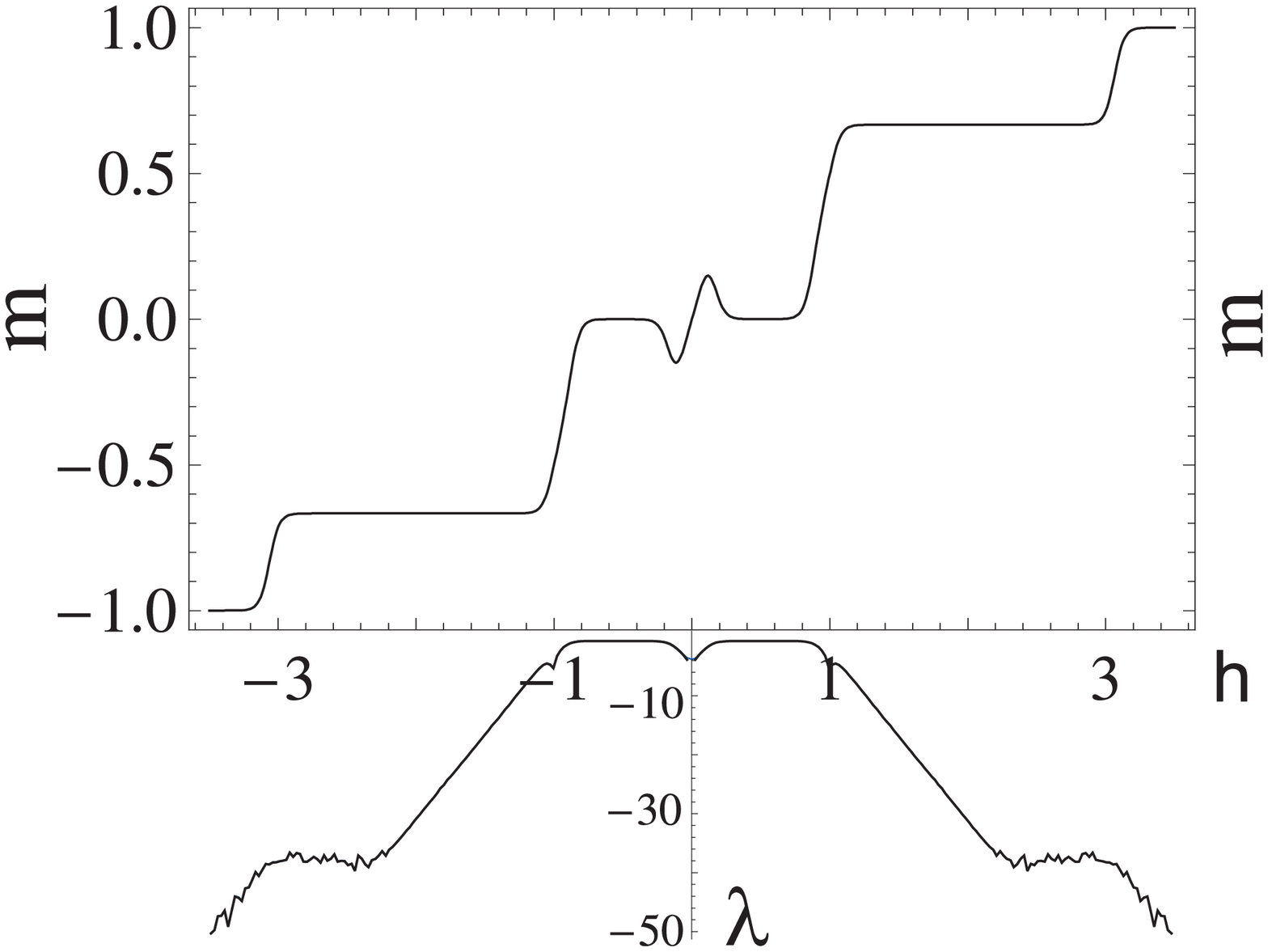}
\caption{Coincidence of magnetization plateaus and the maximum of Lyapunov exponent.}
\end{figure}
Relations (\ref{rec2}) permit the calculation of Lyapunov exponents near the plateaus: in the two-dimensional space they have the form\begin{equation}
\lambda_{1,2}=\lim_{N\to\infty}\frac{1}{N}\ln\left(eigenvalues\{\mathbf{\Lambda}(x_1,y_1)\cdot\mathbf{\Lambda}(x_2,y_2)\cdots\mathbf{\Lambda}(x_N,y_N)\}\right),
\end{equation}
where $\mathbf{\Lambda}(x_i,y_i)$ is Jacobian matrix evaluated at the $(x_i,y_i)$ point
\begin{equation}
\mathbf{\Lambda}(x_i,x_i)=\left. \begin{pmatrix} \frac{\partial\Phi}{\partial x} \ & \frac{\partial\Phi}{\partial y} \\
\frac{\partial\Psi}{\partial x} & \frac{\partial\Psi}{\partial y} \\
\end{pmatrix}\right|_{x_i,y_i}.
\end{equation}

The only subtle point in the estimation of Lyapunov  exponents \cite{Sprott,skokos,kuznecov}
is in the rescaling procedure in each iteration to avoid the numerical
overflow. The results of calculation are shown in Fig. 4 with the corresponding magnetization
function: it is seen in the figures that the maximum Lyapunov exponent also exhibits plateaus. Moreover, for the maximum Lyapunov exponent the location of magnetization plateaus coincide with those of magnetization curves.

\section*{Conclusions }
In the present paper the theory of dynamical systems has been used to study magnetization on the kagome chain. In a strong external magnetic field the Ising model has been considered instead of the Heisenberg one for the case of two, three and six-site exchange interactions. Exact recursion relations have been derived for the partition function. The magnetization curves for different temperatures and exchange parameters were obtained and the kagome chain was observed to split into four sublattices with different magnetizations. Two of them exhibit plateaus at zero and 2/3 of the saturation field value at low temperatures and certain exchange parameters. A multidimensional mapping was deduced to study the maximum Lyapunov exponent near the plateaus and it was shown that the location of plateaus on the maximal Lyapunov exponent and for magnetization was the same.

\section*{Acknowledgement}
This work was supported by the
ANSEF 1518-PS, 1981-PS  and ECSP-09-08\_sasp NFSAT  research grants.


\end{document}